\documentclass[aip,pop,reprint]{revtex4-2}
\usepackage{dcolumn}
\usepackage{amsmath}
\usepackage{bm}
\usepackage{hyperref}
\usepackage{natbib}
\usepackage{amsfonts}
\usepackage{booktabs}
\usepackage{siunitx}
\usepackage{relsize}
\usepackage{mathtools}
\usepackage[colorinlistoftodos]{todonotes}
\usepackage{physics}
\usepackage{graphicx}
\usepackage{array}
\usepackage{float}
\usepackage{svg}
\usepackage{tikz}
\usetikzlibrary{tikzmark}
\usepackage{empheq}
\usepackage{enumitem}
\makeatletter
\def\namedlabel#1#2{\begingroup
    #2%
    \def\@currentlabel{#2}%
    \phantomsection\label{#1}\endgroup
}

 \usepackage{caption}
\makeatother
\usepackage{tcolorbox}
\usepackage{color}
\usepackage{amssymb}

\begin{document}

\title{Simulations of ion heating due to ion-acoustic instabilities in presheaths 
}

\author{Lucas P. Beving}
\affiliation{Applied Physics Program, University of Michigan, Ann Arbor, MI 48109, USA}
\author{Matthew M. Hopkins}
\affiliation{Applied Optical and Plasma Sciences, Sandia National Laboratories, Albuquerque, NM 87185, USA} 

\author{Scott D.\ Baalrud}
\affiliation{Applied Physics Program, University of Michigan, Ann Arbor, MI 48109, USA}
\affiliation{Department of Nuclear Engineering and Radiological Sciences, University of Michigan, Ann Arbor, MI 48109, USA}

\date{\today}

%=======================================================================
%============================ Abstract =================================
%=======================================================================
\begin{abstract}

Particle-in-cell direct simulation Monte Carlo simulations reveal that ion-acoustic instabilities excited in presheaths can cause significant ion heating. 
Ion-acoustic instabilities are excited by the ion flow toward a sheath when the neutral gas pressure is small enough and the electron temperature is large enough. 
A series of 1D simulations were conducted in which neutral plasma (electrons and ions) was uniformly sourced with an ion temperature of 0.026 eV and different electron temperatures (0.1 - 50 eV).
Ion heating was observed when the electron-to-ion temperature ratio exceeded the minimum value predicted by linear response theory to excite ion-acoustic instabilities at the sheath edge ($T_e/T_i\approx28$).  
When this threshold was exceeded, the temperature equilibriation rate between ions and electrons rapidly increased near the sheath so that the local temperature ratio did not significantly exceed the threshold for instability. 
This resulted in significant ion heating near the sheath edge, which also extended back into the bulk plasma; presumably due to wave reflection from the sheath. 
This ion-acoustic wave heating mechanism was found to decrease for higher neutral pressures, where ion-neutral collisions damp the ion-acoustic waves and ion heating is instead dominated by  inelastic collisions in the presheath. 
\end{abstract}
\maketitle

%=======================================================================
%============================ Introduction =============================
%=======================================================================
\section{Introduction}

A property of low temperature plasmas is that the ion temperature is close to room temperature, the temperature of the neutral gas ($T_n, T_i \sim 0.026$ eV), while the electron temperature can be much higher ($T_e \sim 0.1-100$ eV). This disparity results from a lack of thermal relaxation between the electrons and ions since the ion-electron mean free path is typically much longer than the device. However, we find that this is not always the case and that the ions can heat significantly without an external heating mechanism.

% A property of low temperature plasmas is that the electrons are much hotter than the ions and neutral gas. 
% This is because the electrons are generally heated by an energy input mechanism and do not cool significantly since the electron-ion and electron-neutral thermal relaxation length scale are much longer than the device.
% % the ion-ion and ion-neutral thermal relaxation length scale is much shorter than the typical device length scale, whereas the electron-ion and electron-neutral thermal relaxation length scale is much longer. 
% The result is that ions and neutral gas are near room temperature ($T_n, T_i \sim 0.026$ eV), while the higher electron temperature ($T_e \sim 0.1-100$ eV) is set by energy input mechanisms, neutral pressure, and device size. 

Here, particle-in-cell (PIC) simulations show that ion heating can result from the natural excitation of ion-acoustic instabilities near plasma boundaries. 
Specifically, ion flow through the presheath can excite ion-acoustic instabilities. 
Scattering from the resulting collective fluctuations acts to rapidly increase the rate of Coulomb collisions \cite{baalrud_kinetic_2008}, and as a consequence, the electron-ion thermal relaxation rate. 
This results in significant ion heating so that the ion temperature takes a value between the neutral and electron temperatures ($T_n \ll T_i \ll T_e$).
Its value at the sheath edge is observed to be set by the threshold (afterword ``th'' superscript) electron-to-ion temperature ratio required for instability $T_e/T_i \approx (T_e/T_i)^{\textrm{th}}$. 
Although the heating is mostly localized to the presheath, ion-acoustic waves reflect from the sheath and can lead to heating in the bulk plasma as well. 

This finding is particularly influential in the context of applications that use a plasma and sheath to create a high emittance ion source, such as in plasma etching of semiconductors or ion beam generation \cite{lieberman_principles_2005,coburn_plasma_nodate,pelletier_plasma-based_2005,sasao_ion_2005}. It may also be applicable to some electric propulsion systems, where ions are accelerated to higher velocities to generate thrust, usually in the presence of hot electrons \cite{takahashi_helicon-type_2019,sanchez-villar_coupled_2021}, and to energy transport controlled by ion-acoustic instabilities in other contexts.~\cite{glanz_ion-acoustic_1981,yee_electron_2017,scheiner_theory_2015,jorns_ion_2014,lafleur_characteristics_2017,hara_overview_2019,kaganovich_physics_2020}

The common expectation is that the ion temperature at the sheath edge is near room temperature at low neutral gas pressures, but can increase at higher pressures due to inelastic collisions (such as ionization and charge exchange) in the presheath \cite{claire_laser-induced-fluorescence_nodate,meige_ion_2007}.
Inelastic collisions cause ions to be sourced at different locations of the presheath potential drop, causing the ion velocity distribution function (IVDF) to broaden, i.e., heat, as the sheath edge is approached. 
As in the previous work of Meige \emph{et al.} \cite{meige_ion_2007}, in our simulations we observe this to be the dominant ion heating mechanism at sufficiently high neutral pressures. 
However, we also find that a fundamentally different instability-enhanced heating mechanism dominates at sufficiently low neutral pressures. 
Ion acoustic instabilities are driven by inverse electron Landau damping \cite{gurnett_introduction_2017}, and therefore require a kinetic description of electrons to capture. 
This is presumably why the previous simulations \cite{meige_ion_2007}, which used a Boltzmann density relation to model electrons, did not see the effect, even at low pressure. 
% hybrid

Experiments using laser-induced fluorescence (LIF) may be able to test the proposed heating mechanism by measuring the IVDF in the presheath. 
%where ion-acoustic instabilities are present. 
Although a number of such measurements have been made \cite{claire_laser-induced-fluorescence_nodate,lee_measurements_2007,yip_verifying_2014}, they mostly focus on plasma conditions where it is difficult to make a definitive test. 
%parameters at which the instability is either not expected or the predicted ion heating is minimal. 
%datasets do not provided a definitive test. 
%only one was made in a plasma parameter regime where the instability-enhanced heating would be expected to be significant [ref]. %Furthermore, none confirmed the presence of ion-acoustic instabilities.
For example, the measurements of Claire \emph{et al.}~\cite{claire_laser-induced-fluorescence_nodate} were made in an argon plasma with an electron temperature high enough to be expected to excite instability ($T_e = 2.5$ eV), but at a high enough neutral gas pressure ($p_n = 0.36$ mTorr) that the associated instability-enhanced ion heating (a factor of 2-3 at this $T_e$) is comparable to what is expected from inelastic collisions with neutrals. 
Similar measurements were made by Lee \emph{et al.}~\cite{lee_measurements_2007} in a xenon plasma, but with a low enough electron temperature ($T_e = 0.61$ eV) that the instability may not have been excited. 
The most pertinent measurements are those by Yip \emph{et al.}~\cite{yip_verifying_2014} made at a low neutral pressure ($p_n = 0.08$ mTorr) and high enough electron temperature ($T_e = 2.4$ eV). 
Indeed, these seem to show evidence of ion heating in the presheath. 
As will be shown in Sec.~\ref{sec:lp}, the measured ion temperature profile throughout the presheath agrees well with our simulations; including a factor of approximately 4 heating of ions near the sheath edge (compared to room temperature).  
Although the existing experiments are consistent with the simulations, they were not designed to test the proposed instability-enhanced ion heating.
An ideal test would vary the electron temperature across its threshold value in a system where the neutral pressure remained sufficiently small ($p_n\lesssim0.1$ mTorr). Such a test would demonstrate whether the instabilities and associated ion heating transition at the predicted threshold of the unstable regime.  
This motivates future experiments to explore lower neutral pressures and higher electron temperatures. 

The existence of ion-acoustic instabilities in the presheath has been discussed in a number of recent works. 
One of the first theories suggested that these instabilities can rapidly thermalize electrons and may contribute to resolving Langmuir's paradox \cite{baalrud_instability-enhanced_2009}. 
Shortly afterward, it was suggested that they also rapidly thermalize ions near the sheath edge as inelastic collisions cause the IVDF to become distorted away from Maxwellian in the presheath \cite{baalrud_kinetic_2011}. 
This latter prediction was tested experimentally by Yip \emph{et al.} ~\cite{yip_verifying_2014} using LIF measurements. 
These measurements showed good agreement between the degree of thermalization and the predicted instability boundary as neutral pressure was varied. 
However, they did not directly probe the instability. 
The first direct measurements of ion-acoustic instabilities in the presheath were provided by recent LIF measurements that leveraged advancements in single photon counting and large collection optics~\cite{hood_laser-induced_2020}.  
These confirmed that ion-acoustic instabilities are present in the presheath and also suggested that the excited waves reflect from the sheath, but did not investigate ion heating. 

Here, we show results from tests of the predicted instability heating using particle-in-cell, direct simulation Monte Carlo (PIC-DSMC) simulations. 
Our 1D-3V simulations applied absorbing boundary conditions on both the left and right boundary causing sheaths to form. 
Ions and electrons were sourced uniformly and at equal rates throughout the domain, with the ions sourced at room temperature ($T^s_i = 0.026$ eV) and the electrons sourced at temperatures ranging from $T^s_e = 0.1-100$ eV. 
To better assess simulation results (especially noise levels), simulations of uniform plasmas in a specular (reflecting) box with similar plasma parameters were also conducted for comparison. 
%while also fixing the potential to 0V at the walls. 
The ion temperature was computed from time-averaged IVDFs and the presence of instabilities was examined by comparing the fraction of energy stored in fluctuations that were ion-acoustic in a nominally stable sheath configuration, an unstable sheath configuration, and a uniform configuration. The energy stored in fluctuations was computed from fluctuations in the charge density.

Results show that instabilities are excited in the presheath, but not in the uniform plasma. 
%, and agree well with the predicted temperature ratio threshold. 
The spectral energy density indicates that a significant fraction of the wave power is reflected from the presheath, in agreement with the suggestion made by Hood \emph{et al.}~\cite{hood_laser-induced_2020} based on experimental measurements. 
Ion heating was also observed to be associated with ion-acoustic instabilities. 
No heating was observed at conditions where instabilities were expected to be absent based on linear theory, but heating near the sheath edge was so significant in the presence of instabilities that the temperature ratio could not significantly exceed the threshold condition for instability $T_e/T_i \approx (T_e/T_i)^{\textrm{th}}$. 
This led to a ``stiff'' boundary condition where the electron-to-ion temperature ratio was locked to the threshold value at the sheath edge. 
For large electron temperatures ($T_e \gtrsim 10$ eV), this led to ion temperatures more than an order of magnitude above room temperature near the sheath edge. 
This rapid ion heating was found to be associated with a significant increase in the electron-ion energy relaxation rate associated with instability-enhanced collisions. This was confirmed by computing the residual of the steady-state ion energy balance equation, which corresponds to the energy moment of the collision operator.
Spatial profiles of this term show that the heating is greatest near the sheath edge, but also extends into the bulk plasma. 
At sufficiently high neutral pressure and low electron temperature, the instability was not present and ion heating was instead found to be controlled by inelastic (ionization) collisions in the presheath.

We also find that the instability-enhanced heating results in a drastically different distribution of ion energy parallel or perpendicular to the single spatial dimension of the simulations. 
Since the waves are confined to a single spatial dimension, the associated instability-enhanced relaxation appears to be as well. 
This leads to highly anisotropic IVDFs with a much larger parallel temperature than perpendicular temperature.

%=======================================================================
%=================== Code, setup, steady-state ==================
%=======================================================================

\section{Simulation setup}

Simulations were conducted using the electrostatic PIC-DSMC code Aleph \cite{timko_why_2012}. 
They used a 1D domain in space, 3D domain in velocity phase-space, and a uniform mesh. A presheath was simulated by applying absorbing boundary conditions at each domain boundary and continuously sourcing electrons and ions uniformly throughout the domain from Maxwellian distributions with respective temperatures $T^s_e$ and $T^s_i$.  Although, this introduces particles into the sheath, such events are expected to be rare and nearly all ions loaded directly into the sheath will not enter the presheath. A uniform source approximates plasma generation in multidipole devices \cite{claire_laser-induced-fluorescence_nodate,lee_measurements_2007,yip_verifying_2014,hood_laser-induced_2020}, where the primary electrons responsible for ionization are expected to have a close to uniform density.

Reference to the electron source temperature ($T^s_e$) will be made often in this work as the independent variable we change to study situations where the ion-acoustic instabilities are or are not present.  However, as a result of losses to the wall the measured electron temperature in a given simulation is slightly lower (at most  20$\%$ lower) than the source temperature; see Tbl.~\ref{tab:Te_ssn}. 
% Aleph assumes
Although they are not simulated, two dimensions perpendicular to the simulated domain are taken to have length 1m when reporting quantities like densities and rate-densities.  This is reflected in Table \ref{tab:Te_ssn} and throughout the remainder of this work.

In each simulation, data were analyzed after the plasma had evolved sufficiently long to reach a steady-state (approximately $600~\mu$s which represents the time needed for a thermal ion to transit the domain), where the number of ions lost to the walls balanced those sourced in the domain. The primary quantity analyzed in steady-state was the ion temperature. Here, ``ion temperature'' refers to the temperature moment calculated from the full 3V velocity distribution function,
\begin{equation}
    T_i = \frac{m_i}{3n_i}\int (\vb{v}-\vb{V}_i)^2f_i(\vb{v})\dd[3]{v},
\end{equation}

\noindent where $m_i$ is the ion mass, $n_i$ is the ion number density, $f_i(\vb{v})$ is the IVDF, and $\vb{V}_i= \frac{1}{n_i} \int \vb{v} f_i(\vb{v}) \dd[3]{v}$ is the ion flow velocity. 
We will later also utilize 1V temperatures from the reduced velocity distribution functions (e.g., $f^x_i(v_x) = \int f_i(\vb{v})\dd v_y \dd v_z$),
\begin{equation}
    T^x_i = \frac{m_i}{n_i}\int (v_x-V_i^x)^2f^x_i(v_x)\dd{v_x},
\end{equation}
\noindent where $V_i^x= \frac{1}{n_i} \int v_x f^x_i(v_x) \dd{v_x}$. We define $T^\parallel_i = T^x_i$ to be the parallel temperature and $T^{\perp}_i = T^y_i$ to be the perpendicular temperature. All simulated quantities were effectively equal in the $y$ and $z$ directions since there was no spatial variation in either direction. The 3V temperature can be computed from the 1V temperatures $T_i = (T^x_i+T^y_i+T^z_i)/3 = (T^{\parallel}_i/3+2T^{\perp}_i/3)$.

Of specific importance is the ion temperature measured at the sheath edge.  Here, we define the sheath edge as the position where the charge density first exceeds $5\%$ of the average ion density.  This value was chosen since it was the smallest value we could resolve with the data from each simulation.  The sheath edge position is nearly the same in each of the simulations in Table \ref{tab:Te_ssn} at 0.496 m.

\setlength{\tabcolsep}{10pt}
\renewcommand{\arraystretch}{1.5}
\begin{table}
  \begin{tabular}{c||c|c|c}\toprule
    $T^s_e (eV)$ & $T_e (eV)$ & $n_0 (\#/m^{3})$ & $R^s (\#/m^{3}/s)$\\\toprule
    0.1 & 0.08 &$1\times10^{13}$& $3.1\times10^{16}$\\
    0.2 & 0.16 &$2\times10^{13}$& $8.76\times10^{16}$\\
    0.5 & 0.41 &$5\times10^{13}$& $3.47\times10^{17}$\\ 
    0.8 & 0.65 &$8\times10^{13}$& $7.01\times10^{17}$\\
    1.5 & 1.25 &$1.5\times10^{14}$& $1.8\times10^{18}$\\
    3 & 2.58 &$3\times10^{14}$ & $5.09\times10^{18}$\\
    6 & 5.40 &$6\times10^{14}$ & $1.44\times10^{19}$\\
    12 & 11.50 &$1.2\times10^{15}$& $4.08\times10^{19}$ \\
    24 & 25.10 &$2.4\times10^{15}$ & $1.15\times10^{20}$\\
    48 & 42.80 &$ 4.8\times10^{15}$ & $3.26\times10^{20}$\\ \bottomrule
  \end{tabular}
  \caption{Simulation parameters. Here $T_e$ is the simulated electron temperature in the middle of the domain. The simulation density was chosen to keep $dx/\lambda_{De} = 0.42$ constant while the electron source temperature was increased. $R^s$ is the rate-density at which electrons and ions were sourced in the simulation. In section III B, simulations A and B refer to the rows where $T_e^s = 6$ eV and $T_e^s = 0.1$ eV respectively.
  }
  \label{tab:Te_ssn}
\end{table}

The 1D spatial grid was composed of $1600$ cells of length $dx = 3.125\times10^{-4}$ m, resulting in a total domain length of $L=0.5$ m, which ensured that the Debye length was resolved. The steady-state density was changed (via the source rate) along with $T^s_e$ to maintain the same electron Debye length ($\lambda_{De} = 7.43\times 10^{-4}$ m). This was done so that the same spatial grid could be used in each simulation. In an experimental test of the instability heating mechanism such control of the plasma density would not be necessary. The steady-state densities of each simulation are summarized in Table \ref{tab:Te_ssn} The average number of computational particles per cell in the center of each simulation was 30 for each species and decreased to 15 near the sheath edge for computational electrons and ions. Fewer computational electrons and ions were found in the sheath; however, this is not expected to affect our measurements in the presheath since ions in the sheath quickly exit the simulations. For simulations with $T^s_e < 24$~eV, a time step of $1.1\times10^{-10}$ s was chosen to meet a CFL-like condition, so that an electron with velocity lower than $2v_{Te} = \sqrt{8k_B T_e/m_e}$ does not cross an entire spatial cell in one time step \cite{courant_partial_1967}. 
For $T^s_e > 24$~eV, a time step of $2.5\times10^{-11}$ s was chosen.

The plasma was assumed to be generated from pure helium, and to be singly ionized. 
Plasma parameters were chosen to match those in the low-temperature plasma experiments commonly used to study sheaths: $T^s_i=0.026$ eV, $n \approx 5\times10^{15}$ m$^{-3}$.\cite{hood_laser-induced_2020,yip_verifying_2014,claire_laser-induced-fluorescence_nodate}
The neutral pressure was varied in order to study different regimes of ion-acoustic wave damping. 
Ion-neutral and electron-neutral elastic collisions were modeled by way of the DSMC method\cite{bird_molecular_1994}. Cross sections for both interactions were from the Phelps database provided by LXCat \cite{noauthor_phelps_nodate-1,noauthor_phelps_nodate,crompton_momentum_1967,hayashi_recommended_1981}.
No explicit method, like DSMC, was used to model Coulomb collisions in the simulations.  However, Coulomb collisions occurring over large enough distances to be resolved by the grid are expected to be included implicitly in the simulations.

%=======================================================================
%======================= Results  ======================================
%=======================================================================
\section{Results at low pressure\label{sec:lp}}
\subsection{Ion heating}
\begin{figure}
    \includegraphics[width=8cm]{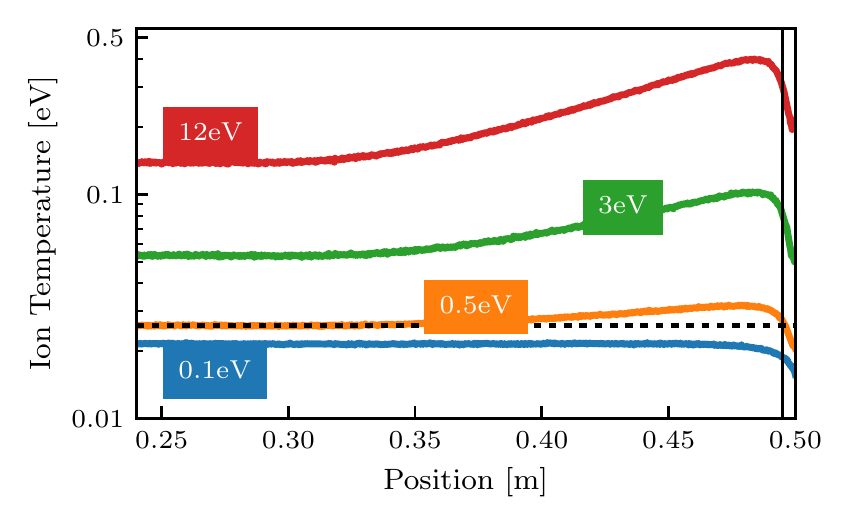}
    \caption{Spatial distribution of ion temperature from simulations with a neutral pressure $p=0.01$ mTorr and four values of the electron source temperature ($T^s_e$) indicated in the boxes. The solid vertical line represents the sheath edge, while the horizontal dashed line represents the ion source temperature $T^s_i=0.026$ eV.}
    \label{fig:Ti}
\end{figure}

Figure~\ref{fig:Ti} illustrates the most notable result, which is the observation of ion heating near the sheath edge (vertical line) at low pressure ($p_n = 0.01\ \text{mTorr}$) when the source electron temperature was sufficiently high.  
The heating is mostly localized to near the sheath edge, but the ion temperature in the bulk plasma is also well above room temperature.
%We attribute this effect to the presence of ion-acoustic instabilities excited near the sheath edge. 
Ion heating is surprising because the only energy input is the kinetic energy at which the electrons and ions are sourced. 
Ion heating to above room temperature requires a mechanism for energy transfer from electrons to ions. 
However, the collisional electron-ion energy relaxation rate based on Coulomb collisions is expected to be negligible. 
Using an estimate based on the standard Coulomb collision frequency \cite{huba_nrl_1998}, the mean free path for electron-ion energy equilibration is thousands of meters at these conditions. Furthermore, since the simulations do not include an explicit Coulomb collision model only long-range interactions resolved by the grid will be included. 
Since there are only about 2 grid points per Debye length, Coulomb collisions that are implicitly simulated are expected to be extremely rare (in the absence of instability). 
%lowering the effective collision frequency even lower. 
% Furthermore, since the simulations do not include an explicit Coulomb collision model, the electron-ion collision frequency is expected to be effectively zero. 

Some ion heating near the sheath edge is expected due to the interaction between the flowing ion distribution and the stationary background neutrals~\cite{meige_ion_2007}. 
However, this effect is negligible at the low pressure of these simulations since the estimated ion-neutral collision mean free path is approximately 10 m. It also does not contribute to heating in the bulk plasma. Alternatively, the time averaged IVDF could broaden (i.e. heat) if the IVDF were to oscillate in velocity space on a shorter time scale than the average.  We find that this is not a significant source of heating in our simulations when we compare snapshots of the IVDF that resolve the ion plasma period.
% Similarly, some heating is expected due to the fact that ions are sourced at different values of the plasma potential throughout the presheath; as in the classical Tonks-Langmuir model \cite{tonks_general_1929}. 
% However, again, this cannot explain heating in the center of the domain. 
% It also does not explain the observed kinetic energy balance of ions through the domain; as will be explained in Sec.~\ref{sec:e_balance}. 

%\textcolor{blue}{It would be great to include a statement here that explains why the observed data cannot be explained by the uniform ion sourcing in the domain. } 

The simulation data appear to indicate that the source of ion heating is enhanced electron-ion energy exchange resulting from ion-acoustic instabilities near the sheath edge, which extends into the plasma due to wave reflection from the sheath. 
A number of observations provide evidence for this. 

First, the parameters at which heating is observed seem to correspond well with the conditions at which the linear dispersion relation for ion-acoustic instabilities predicts instability in the presheath. 
The threshold for exciting the instability can be estimated from the ion-acoustic dispersion relation, which has real frequency
\begin{equation}
    \omega_r = kV_i^x - \frac{kc_s}{\sqrt{1+k^2\lambda_{De}^2}}
    \label{eq:wr}
\end{equation}
and growth rate \cite{gurnett_introduction_2017}
\begin{align}
   \gamma = & \frac{-kc_s\sqrt{\pi/8}}{\qty(1+k^2\lambda_{De}^2)^{2}}\Biggl[\qty(\frac{T_e}{T_i})^{3/2}\exp\left(-\frac{T_e/T_i}{2(1+k^2\lambda_{De}^2)}\right) \Biggr. \nonumber\\
   +&\Biggl.\sqrt{\frac{m_e}{m_i}}\biggl(1-\frac{V_i^x}{c_s}\sqrt{1+k^2\lambda_{De}^2}\biggr)\Biggr].
    \label{eq:gamma}
\end{align}
Here, $k$ is the wavenumber, $c_s=\sqrt{k_B T_e/m_i}$ is the ion sound speed, and $\lambda_{De} = \sqrt{\epsilon_o k_BT_e/e^2 n_e}$ is the electron Debye length. The change in density between simulations with different electron source temperatures does not affect this prediction since the density only appears in the Debye length, which is held constant. The threshold temperature ratio decreases with increasing flow velocity and reaches a value of $(T_e/T_i)^\textrm{th} \approx 28$ when the flow speed reaches its maximum value in the presheath, which is approximately the sound speed. This estimation corresponds to $T^\textrm{th}_e = 0.7$~eV if the ions have the source temperature, $T_i = 0.026$~eV. In a plasma with heavier ions the threshold temperature ratio increases slightly: 30 for argon, 30.3 for xenon, and 30.8 for krypton.

%The instability boundary can be estimated the zero contour of the maximum growth rate and is shown as the solid black line in Figure \ref{fig:inst_bndry}. 
%Since $(T_e/T_i)^{th}$ is the minimum temperature ratio require to excite the ion-acoustic instability and we expect that $V_i^x$ will not exceed $c_s$ significantly in the presheath we estimate $(T_e/T_i)^{th}$ from the intersection of $V_i^x/c_s = 1$ and the instability boundary.  This estimation results in $(T_e/T_i)^{th} \approx 28$ and $T^{th}_e = 0.7$~eV when we assume $T_i = 0.026$~eV.

Of the four conditions shown in Fig.~\ref{fig:Ti}, the two cases with source electron temperatures above the predicted threshold ($T^s_e = 3$~eV and $T^s_e = 12$~eV) both clearly exhibit significant heating.
No heating is observed in the case with source electron temperature well below the threshold ($T^s_e = 0.1$~eV), though some cooling is observed as a result of losses to the walls. 
The remaining simulation has a source temperature close to the threshold ($T^s_e = 0.5$~eV) and exhibits minor heating near the sheath edge.  The heating may be the result of instabilities since the source temperature is close to the estimated threshold value.

% this case shows no significant heating (there is evidence of slight heating near the sheath edge, but no heating in the rest of the domain). 

\begin{figure}
    \includegraphics[width=\linewidth]{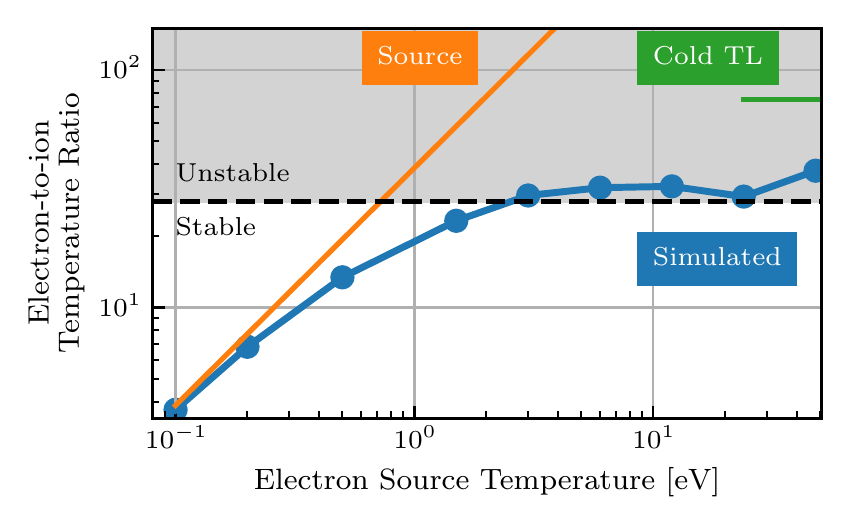}
    \caption{Observed electron-to-ion temperature ratio at the sheath edge (blue circles) as the electron source temperature ($T^s_e$) is varied. The dashed line represents the predicted threshold for instability from Eq.~(\ref{eq:gamma}), and the orange line represents the source temperature ratio ($T^s_e/T^s_i$). The green line represents the temperature ratio at the sheath edge predicted by the Tonks-Langmuir model.}
    \label{fig:Tr}
\end{figure}

Second, if the source temperature ratio is high enough to expect instability in the presheath, the observed temperature ratio at the sheath edge (subscript ``se'') takes a value approximately equal to the instability threshold: $(T_e/T_i)_{\textrm{se}} \approx (T_e/T_i)^{\textrm{th}}$; as depicted by the horizontal dashed line in Fig.~\ref{fig:Tr}.  Here, the orange line represents the source temperature ratio $T^s_e/T^s_i$, which would also be the expected temperature ratio at the sheath edge in the absence of any ion heating (or cooling).  In addition, some ion heating is expected in the presheath due to the fact that ions are sourced at different values of the plasma potential throughout the presheath; as in the classical Tonks-Langmuir model \cite{tonks_general_1929}. The green line represents the sheath edge temperature ratio predicted by the Tonks-Langmuir (TL) model, where $(T_e/T_i)_{\textrm{se}}^{TL} = 3(1/25+2T^s_i/T_e)^{-1}\approx 75$ was used to convert the 1V TL prediction to the 3V of our simulations.  Since the ion source is cold in the TL model, its prediction only applies to simulations where $T^s_e/T^s_i>>1$.
It is important to note that the TL model, even with a warm ion source, does not predict heating in the center of the domain \cite{sheridan_solution_2001} and does not explain the observed kinetic energy balance of ions through the domain; as will be explained in Sec.~\ref{sec:e_balance}.

Instead of following the orange line, or locking to the green line, the observed temperature ratio at the sheath edge (blue dots) locks to the threshold value, $(T_e/T_i)^{\textrm{th}} \approx 28$, when the source electron temperature ($T^s_e$) exceeds the instability threshold ($T_e^{\textrm{th}} \approx 0.7$~eV).  
This implies that the plasma cannot significantly enter an unstable parameter regime. 

\begin{figure}
    \includegraphics[width=\linewidth]{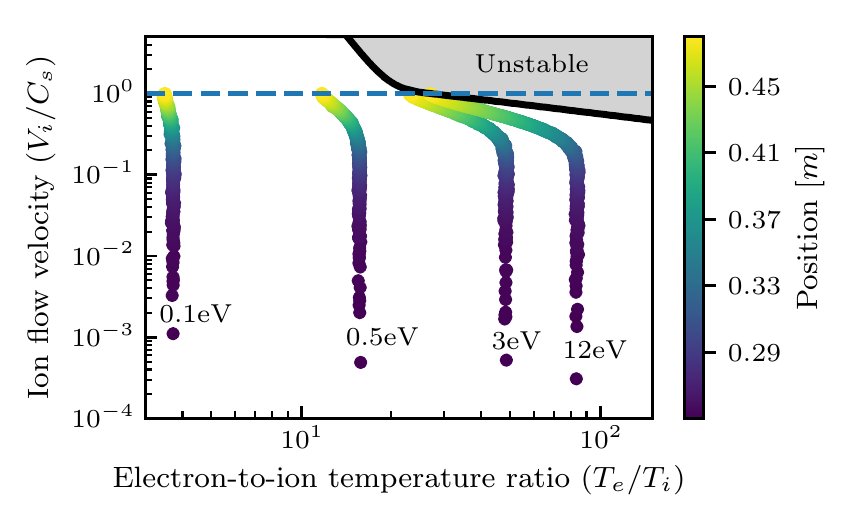}
    \caption{Ion flow speed ($V_i^x/c_s$) and electron-to-ion temperature ratio ($T_e/T_i$) through the presheath from simulations with varying electron source temperatures ($T^s_e$) indicated by the numbers in the figure. The dashed line indicates the sheath edge, where the Bohm criterion is met, $V_i^x/c_s = 1$, and the shaded region indicates parameters predicted to be ion-acoustic unstable according to Eq.~(\ref{eq:gamma}). }
    \label{fig:inst_bndry}
\end{figure}

Third, when instability is predicted, the observed ion heating extends far enough into the presheath that the plasma does not significantly enter the unstable parameter regime. 
Fig.~\ref{fig:inst_bndry} shows the ion flow velocity and temperature ratio as a function of position in each of the four simulations from Fig.~\ref{fig:Ti}. 
%the data from each simulation is represented by a different curve labeled by $T^s_e$ and colored by the position. 
In each simulation the ions flow toward the sheath edge with increasing velocity until they satisfy the Bohm criterion with $V_i^x/c_s = 1$. 
The lines are expected to be purely vertical in the absence of ion (or electron) heating. 
As expected, a nearly vertical line is observed when the electron source temperature is in the stable region ($T^s_e =0.1$~eV). 
%simulation forms a nearly straight line as the ions are accelerated up to $c_s$ and the temperature ratio remains effectively constant since there is no heating. 
However, when the electron source temperature is high enough that isothermal ions would lead to instability near the sheath edge ($T^s_e = 3~\text{eV}$ and $T^s_e = 12~\text{eV}$), the ions are observed to heat in the presheath well before reaching the instability boundary. 
Ions reach the instability boundary at the sheath edge, where the temperature ratio takes the universal value associated with the intersection of the Bohm criterion ($V_i^x=c_s$) and the instability threshold. 
The fact that the ions heat further into the presheath than the location at which instability is predicted suggests that the excited ion-acoustic waves may reflect from the sheath and propagate back through the presheath and into the bulk plasma. 
Further evidence for this will be shown in Sec. \ref{sec:instability}. 
%simulations the ions approach the unstable region (grey) as they accelerate toward the sheath and cannot slow down as they must reach the sound speed.  Since they cannot slow, the instability is activated and the ions heat. 
%The result is that the plasma does not continue along a vertical line into the unstable region, but sits at the intersection of the instability boundary and $V_i^x/c_s = 1$. In each simulation with $T^s_e >T^{th}_e$ the ions heat so that the sheath edge temperature ratio becomes $\approx 28$.

\begin{figure*}
%\centering
\includegraphics[width=\textwidth]{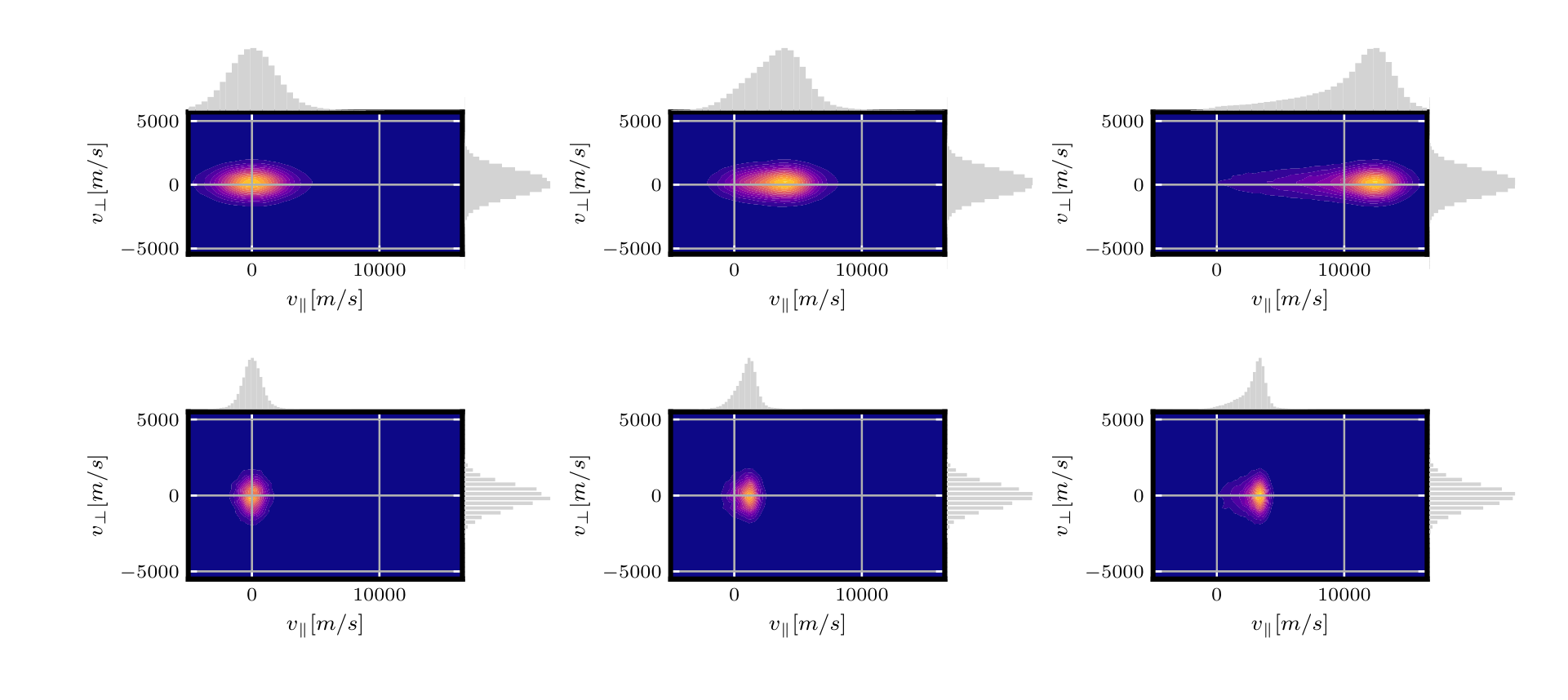}
\caption{IVDFs in the $v_{\perp}$-$v_{\parallel}$  ($v_{y}$-$v_x$) plane from simulations with a high electron source temperature ($T^s_e=6$~eV, top row) and a low electron source temperature  ($T^s_e=0.5$~eV, bottom row).  Each is plotted at three different locations from left to right: middle of the domain ($0.25$~m), presheath ($0.37$~m), and sheath edge ($0.49$~m).}
\label{fig:lp_IVDFs}
\end{figure*}

\begin{figure}[h]
    \centering
    \includegraphics[width=\linewidth]{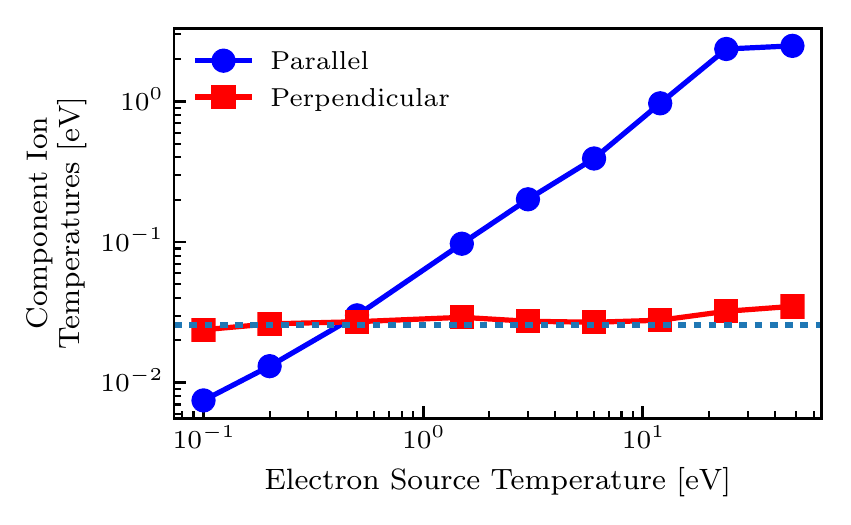}
    \caption{Parallel and perpendicular components of the ion temperature as a function of the electron source temperature ($T^s_e$). The dashed line indicates the ion source temperature ($T_i^s = 0.026$~eV).}
    \label{fig:Tperp}
\end{figure}

The observed ion heating leads to a significant temperature anisotropy in the presheath. 
In particular, Figs.~\ref{fig:lp_IVDFs} and \ref{fig:Tperp} show that the ion heating is entirely constrained to the single dimension of the spatial domain. 
The IVDFs shown in Fig.~\ref{fig:lp_IVDFs} illustrate that the ion temperature remains isotropic (or nearly so) throughout the presheath in the case where instability is predicted to be absent (see the $T_e^s = 0.5$ eV case). 
In contrast, the IVDF spreads significantly in the parallel dimension when instability is predicted ($T_e^s = 6$ eV).  It is noteworthy that the ion-acoustic group velocity lies below the peak of the IVDF, on the low-velocity tail, and is the region of velocity phase-space at which wave-particle scattering is expected to be most frequent. 
In each of the 6 panels of Fig.~\ref{fig:lp_IVDFs}, the reduced perpendicular IVDFs are nearly the same, demonstrating that the perpendicular IVDF is not affected. 
Additionally, the perpendicular IVDF does not evolve significantly throughout the presheath. 
These observations are consistent with wave-particle scattering, which would be expected to be confined to the single dimension in which the wave exists in these 1D simulations. The fact that the simulations are 1D-3V may preclude any affects to the perpendicular IVDF by the waves, which would be expected to spread in 3 dimensions in reality. 
Studying this will require simulations in 2D or 3D. 

Aggregate data showing the parallel and perpendicular ion temperatures at the sheath edge are shown in Fig.~\ref{fig:Tperp}. This shows a consistent story with the IVDFs, demonstrating that the ion heating is constrained to the parallel direction.  The reason the parallel temperature dips below the source temperature for low $T^s_e$ is because the boundaries provide an energy sink for ions that does not exist in the perpendicular directions. 
In the absence of the ion-ion Coulomb collisions required to thermalize the distribution, an anisotopy forms with a higher perpendicular temperature.
%some energy is lost with the particles that exit at the boundaries. 
This effect is not noticeable at higher $T^s_e$ where the instability heating becomes significant and dominates the cooling effect.

Finally, we add that these simulation results compare well with the LIF measurements of Yip \emph{et al.}~\cite{yip_verifying_2014}. Even though the experiments used a xenon plasma the threshold electron-to-ion temperature increases only slightly to 30.3. This means that for a given electron temperature (e.g., 3 eV) we should expect the ion temperature at the sheath edge to decrease from 0.107 eV in a helium plasma to 0.099 eV in xenon.
In particular, compare the simulation using $T_e^s=3$~eV from Fig.~\ref{fig:Ti} with the measured ion temperature profile in Fig.~5 from Ref.~\onlinecite{yip_verifying_2014}, which was obtained in a plasma with an electron temperature of $T_e \approx 2.4$~eV and a neutral pressure of $0.08$~mTorr. 
The electron temperature in this comparison is close enough that we expect similar results for the ion temperature profile, and the neutral pressure in the experiment is low enough that we expect the instability to not be affected by neutral damping. 
The ion temperature profile in the simulation is consistent with that measured in the experiment. 
In both, the ion temperature near the sheath edge is $0.1$~eV, which is significantly higher than the room temperature ion source. 
Similarly, in the experiment the ion temperature reported farthest from the sheath was $\approx0.05$~eV, which is close to the value of $0.053$~eV observed at the center of the domain in our simulation.

\subsection{Ion-acoustic instabilities \label{sec:instability}}

\begin{figure*}
    \centering
    \includegraphics[width=\textwidth]{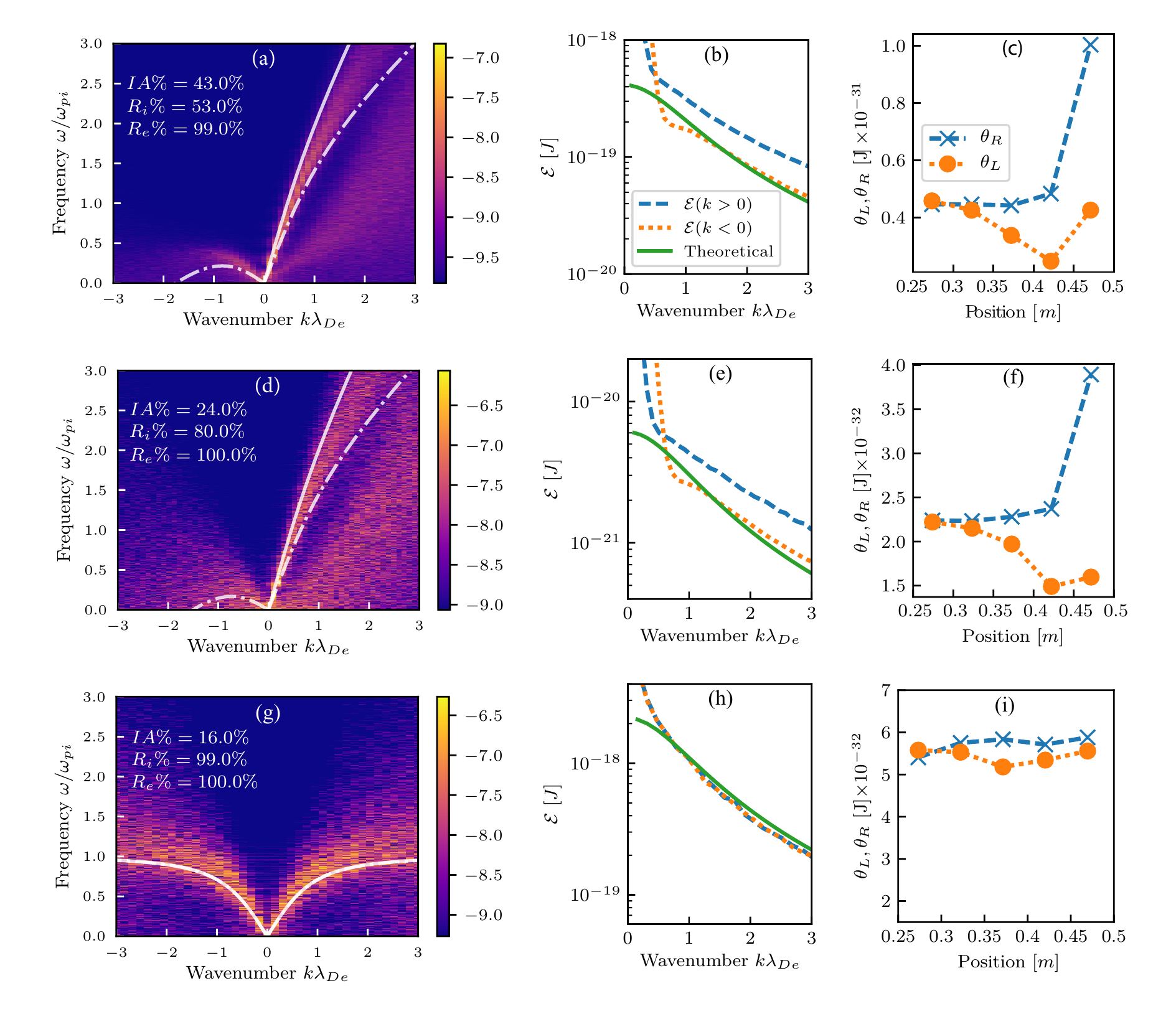}
    \caption{Fourier transform of ion density fluctuations ($\log_{10}((\delta n_i/n_i)(k,\omega))$) in simulations A [$T^s_e = 6$~eV: (a)], B [$T^s_e = 0.1$~eV: (d)], and C [uniform: (g)]. White dashed and solid lines indicate the real part of the ion-acoustic dispersion relation (Eq.~(\ref{eq:wr})) calculated using data at 0.447 m and 0.496 m, respectively. $I\!A\%$ denotes the percent of the signal that is ion-acoustic, $R_i\%$ is the reflection coefficient of wave energy carried by ions from the sheath, and $R_e\%$ is that for electrons. Panels (b), (e), and (h) show the spectral energy between 0.447 m and 0.496 m calculated using Eq.~(\ref{eq:spec_energy}) for $k>0$ (dashed) and $k<0$ (dotted) in simulations A,B, and C. The solid line represents the theoretical prediction from Eq.~(\ref{eq:theory_spec_energy}). (c,f,i) show profiles of the energy per particle carried by right- (crosses) and left- (circles) moving ion-acoustic modes in simulations A, B, and C}
    \label{fig:density_flucs}
\end{figure*}

The presence of instabilities are expected to enhance the electrostatic fluctuation level above the thermal level. To investigate this, we compare the fraction of fluctuations that are ion-acoustic (those with $k$ and $\omega$ near that of Eq.~(\ref{eq:wr})) from a simulation where the ion acoustic instabilities are expected (simulation A: $T^s_e = 6$~eV) to one where they are not (simulation B: $T^s_e = 0.1$~eV); see Fig.~\ref{fig:density_flucs}.  
The results show a higher fraction of ion-acoustic fluctuations in case A, which matches the expectation that ion-acoustic instabilities enhance coherent ion-acoustic fluctuations.  
Furthermore, the sheath simulations are compared with a simulation of a uniform plasma (simulation C), which exhibits only thermal fluctuations. 

Ion-acoustic fluctuations are expected even in simulations where no instabilities or presheaths are present.  We designed the uniform simulation to check that the simulated fluctuation level matches the theoretical level for a uniform plasma.  The simulation utilized reflecting boundary conditions for all particles at the walls and fixed the potential to 0 V there as well. The same average number of particles per cell was used in the uniform simulation as in the presheath simulations, so that the statistical noise would be the same. The density and electron temperature were chosen to be $6\times 10^{14} ~\textrm{m}^{-3}$ and $6$ eV. However, numerical heating is noticeable in the simulation since it is a closed system.  The time averaged electron temperature is higher (12 eV) than the initial temperature (6 eV) since the duration of simulation C extends over at least several microseconds (thousands of $\omega_{pe}^{-1}$) to acquire enough data for the analysis.  The acquisition time for the uniform simulation is shorter than the presheath simulations, which results in a more granular image in Fig.~\ref{fig:density_flucs} (g).
% to determine if the level of ion-acoustic fluctuations present in a simulation with no instabilities or presheath and

Fluctuations in each simulation were analyzed using 2D (space and time) Fourier transforms of the ion density (density spectrum) between a point in the presheath ($0.447$ m) and the presheath-sheath boundary ($0.496$ m). Fig.~\ref{fig:density_flucs} shows the logarithm of the ion density spectrum near the right boundary in simulation A (a), B (d), and C (g). The color axis in each plot was set to range from the maximum spectral value down to 100 times less than the maximum since the maximum varies between simulations. We identified ion-acoustic modes in each simulation by comparing the density spectrum to the theoretical dispersion of ion-acoustic modes from Eq.~(\ref{eq:wr}). The white lines denote the real frequency calculated from the average densities, temperatures, and velocities at $0.447$m (dashed) and $0.496$m (solid). We see that in each simulation some of the fluctuations are ion-acoustic since they fall within the predicted dispersion relation curves, while part of the signal does not and is representative of \textit{thermal noise}, defined as in Dieckmann \textit{et. al}.\cite{dieckmann_simulating_2004}

% the inherently large noise in PIC simulations \cite{langdon_kinetic_1979,dieckmann_simulating_2004}.

Ion-acoustic instabilities are expected to increase the ion-acoustic fluctuation level above the thermal noise level represented by the uniform simulation. However, the thermal noise level depends on the electron temperature \cite{langdon_kinetic_1979} and numerical heating makes it difficult to match the electron temperature of simulation C to that of either A or B. 
For this reason, we compare the fraction of the total energy stored in the ion-acoustic fluctuations rather than the absolute fluctuation levels. The fraction of energy that is ion-acoustic was calculated as
\begin{equation}
    I\!A\% = \frac{\int_{IA}\hat{\rho}^2/(2\epsilon_0k^2)\dd{k}\dd{\omega}}{\int \hat{\rho}^2/(2\epsilon_0k^2)\dd{k}\dd{\omega}},
    \label{eq:ia_frac}
\end{equation}
where $\hat{\rho}$ is the charge density spectrum, $\epsilon_0$ is the permitivity of free space, and the integral in the numerator is carried out only over ion-acoustic modes. 
Referencing Fig.~\ref{fig:density_flucs} (a,d,g), the range of the ``$IA$'' integral includes values up to $0.3\omega_{pi}$ greater and less than the frequency range predicted by the linear dispersion relation (solid and dashed white lines). This was done to exclude the part of the signal that clearly does not fall within the ion-acoustic lines (e.g. signal in the bottom right of (a) and (d)).
It includes only positive values of the frequency and does not include any part of the signal with $\abs{k\lambda_{De}} < 0.1$ as this represents the smallest wavenumber that could be resolved. 
The fraction of the total energy that is ion-acoustic is reported in each of the spectrum plots and is noticeably higher in simulation A ($IA\% = 43\%$) compared to B ($24\%$) or C ($16\%$).  
The higher ion-acoustic wave energy fraction in simulation A compared to B is consistent with the prediction of Eq.~(\ref{eq:gamma}) that ion-acoustic instabilities are excited in A but not B.

Fluctuations in the uniform simulation (C), as well as the left moving modes ($k<0$) in both A and B, are consistent with thermal noise.  This can be confirmed by calculating the spectral energy as
\begin{equation}
    \mathcal{E}(k) = V\int \frac{\hat{\rho}^2}{2\epsilon_0k^2}\dd{\omega},
    \label{eq:spec_energy}
\end{equation}
where $V$ is the volume of the Fourier transformed region near the right boundary ($V = 0.049$~m$^3$).
We compare this to the theoretical value of the spectral energy for a uniform and stable plasma~\cite{rostoker_fluctuations_1961}
\begin{equation}
    \mathcal{E}(k)_{\textrm{theory}} =\frac{T_e^c}{2}\frac{1}{1+k^2\lambda_{De}^2}.
    \label{eq:theory_spec_energy}
\end{equation}
Here $T_e^c = wT_e$ is the temperature of the computational electrons, where $w$ is the number of real electrons represented by a computational electron. Although the effects of ion thermal motion can be included in this prediction, the effects are small \cite{rostoker_fluctuations_1961}.
In the uniform plasma (simulation C), both left ($k<0$) and right ($k>0$) moving modes agree well with the theoretical predication for thermal fluctuations; see panel (h), as expected. 
%This is expected since ion-acoustic instabilities are not expected in a uniform plasma. 
In contrast, panels (b) and (e) show that the energy in the right moving modes is higher than the thermal level in both simulations A and B.   
Although the enhanced level of fluctuations beyond the predicted thermal level might be considered evidence of instability, there are other reasons this may occur: The estimate from Eq.~(\ref{eq:theory_spec_energy}) is for a uniform plasma and does not account for gradients in plasma parameters such as density or ion flow, which are both characteristic features of the presheath. 
Thus, the comparison between the observed and thermal spectra does not provide a conclusive test for instability. 

Observation of the instabilities may also be obscured by the fact that the thermal noise level is enhanced in PIC simulations by the particle weight \cite{langdon_kinetic_1979,dieckmann_simulating_2004}, as seen in Eq.~(\ref{eq:theory_spec_energy}).  For example, if the instabilities saturate because of ion trapping, we can estimate the saturation energy density as $\theta_{Sat}\approx T_en_e/36$ \cite{lafleur_theory_2016}. This does not change with the particle weight since the computational particle density is proportional to $w^{-1}$ and the computational particle temperature is proportional to $w$. When evaluated for the plasma parameters of simulation A we find $\theta_{Sat}\approx 1\times10^{-3}\:J/m^{3}$.  Furthermore, integrating Eq.~(\ref{eq:theory_spec_energy}) over all $k$ gives an estimate for the PIC thermal energy density of $\theta_{thermal}\approx T_e^c/(\lambda_{De}\times 1m^2)\approx10 \:J/m^3$, which is $10^4$ times higher than the predicted level for saturation. This demonstrates that in such a situation, the instabilities may be difficult to detect by simply comparing observed and thermal spectra.

% Specifically, the fluctuation level of the instabilities may saturate at a level that is much higher than thermal noise level in a real plasma, but comparable or smaller than the thermal noise level in a PIC simulation. 

Profiles of the energy per particle stored in ion-acoustic modes provide further evidence that the ion-acoustic instabilities are excited in simulation A. The energies per particle are calculate as 
\begin{subequations}
\begin{align}
    \theta_L = & \frac{1}{n_i}\int_{IA,k<0}\mathcal{E}(k)\dd{k}, \\
    \theta_R = & \frac{1}{n_i}\int_{IA,k>0}\mathcal{E}(k)\dd{k},
    \label{eq:epp_lnr}
\end{align}
\end{subequations}
where $n_i$ is the local average ion density. Here $\mathcal{E}$ was calculated as in Eq.~(\ref{eq:spec_energy}), but using the charge density spectra in adjacent regions of volume $V$ from the center to the sheath edge. The energy in the right moving modes (crosses) increases, especially near the sheath edge, in both simulations A and B; see panels (c) and (f). However, the energy stored in right moving ion-acoustic modes increases more across the presheath in simulation A than B (a percent increase of 122$\%$ vs 71$\%$). This is consistent with the presences of instabilities in A and not B. In simulation C the energy stored in both modes is equal and does not vary, as seen in panel (i).  In simulations A and B the left moving energies decrease along the direction of ion flow in the presheath except near the sheath edge. The decrease may be because, as one moves toward the sheath, there is less plasma to the right that can emit left moving fluctuations.  If this is the case, then it is surprising that the energy in the left moving modes does not reach 0 near the sheath. In fact it increases near the sheath edge in both cases. Reflection of the right moving modes by the sheath or presheath density gradient may explain this.

% \textcolor{red}{Maybe the left moving wave profile is due to wave reflection from the density gradient too.}

% This is unexpected since left moving noise should still be generated near the sheath;however, it is possible that the energy is stored in wavelengths with $k<0.1$ we have poor resolution.

% The energy scales in each simulation are different as they depend on the electron temperature of the simulation, which is highest in the uniform simulation (f) and lowest in the stable simulation (d) \cite{rostoker_fluctuations_1961}.

\begin{figure*}
    \centering
    \includegraphics[width=\textwidth]{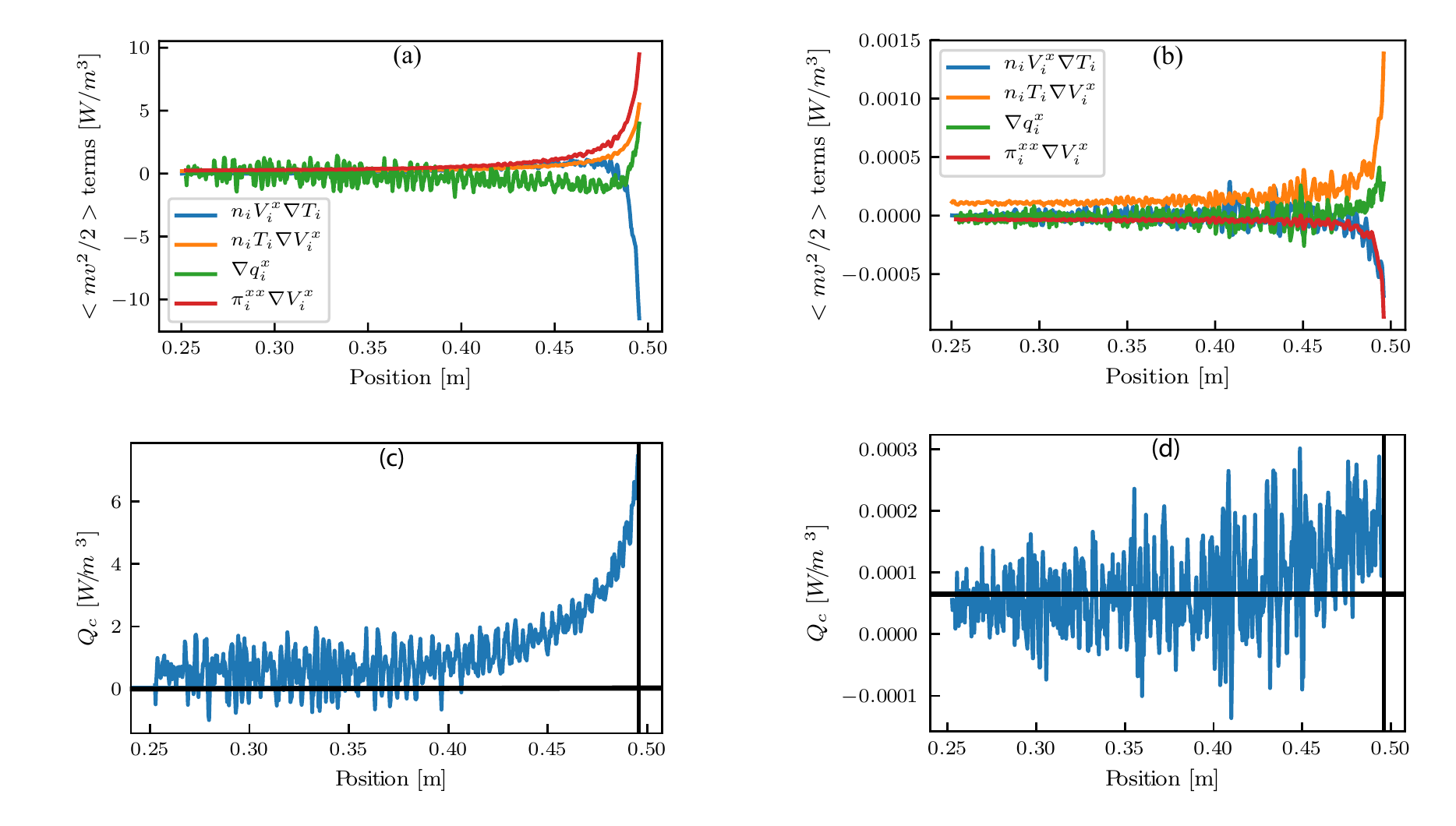}
    \caption{Terms on the left side of the ion energy balance from Eq.~(\ref{eq:1d_temp_evo}) evaluated from the IVDF in: (a) simulation A ($T_e=6$ eV), and (b) simulation B ($T_e=0.1$ eV). The instability-enhanced collision term ($Q_c$: blue line) and the source term ($\frac{3}{2}\nu^s_in_iT^s_i$: horizontal black line) are shown in panel (c) for simulation A and panel (d) for simulation B. The vertical line represents the sheath edge position.}
    \label{fig:Qi_low}
\end{figure*}

We hypothesize that right moving ion-acoustic modes reflect from the sheath and propagate into the bulk plasma. 
%This could explain the heating observed far from the sheath.  
Consider an ion-acoustic wave excited in the presheath that is moving rightward toward the sheath at the sound speed $c_s$. 
Such a wave is carried by electrostatic fluctuations (compression and rarefaction) in both the ions and electrons. 
When the wave reaches the sheath edge, all of the ions are transferred though the sheath to the boundary, carrying the ion fluctuation energy with them. 
However, essentially all of the electrons are reflected. 
The reflected electrons continue to oscillate at the ion-acoustic wave frequency as they propagate back into the bulk plasma. 
Near the sheath, ions are drifting at nearly the ion-acoustic speed to the right, so there are very few ions resonating with the phase speed of the reflected wave propagating to the left. 
Thus, the transfer of wave energy from electrons to ions for negative-$k$ modes is minimal near the sheath. 
However, nearer to the bulk plasma, more ions populate this region of phase space so more fluctuation energy can be transferred from electrons to ions. 
In the bulk plasma, the ion wave energy is observed in both the electrons and ions. 
This picture is consistent with left and right moving wave energy profiles shown in Figs.~\ref{fig:density_flucs}(c) and (f). 
It is also consistent with the observation of ion heating in the bulk plasma though instabilities are only predicted near the sheath edge. 

% Specifically, we propose that the ion-acoustic fluctuations are imprinted on the electrons near the sheath, as they are mobile enough to resolve the low frequency ion-acoustic fluctuations. The electrons then carry the fluctuations back into the presheath when they reflect from the sheath potential. 
% Further analysis of the energy spectrum provides evidence for this mechanism.

To test if the electron fluctuations are reflected from the sheath we calculate the ratio of energy stored in left and right moving electron fluctuations between $0.447$ m and $0.496$ m (approximating the presheath region):
\begin{equation}
    R_{e,i} \% = \frac{\int_{IA,k<0}e^2\hat{n}_{e,i}^2/(2\epsilon_0k^2)\dd{k}\dd{\omega}}{\int_{IA,k>0}e^2\hat{n}_{e,i}^2/(2\epsilon_0k^2)\dd{k}\dd{\omega}}.
    \label{eq:ia_frac}
\end{equation}
Here, $\hat{n}_{e,i}$ is the electron or ion density spectrum.
The values of $R_e$ are shown in Fig.~\ref{fig:density_flucs} (a,d,g), where each is effectively $100\%$.  Such high percentages imply that all the electron fluctuations moving into the sheath are reflected, as we expect.
We test our assumption that ion fluctuations are absorbed at the sheath by computing the same ratio, but using the ion density spectrum in Eq.~(\ref{eq:ia_frac}); providing $R_i$. In both presheath simulations the ion fluctuations are partially absorbed since the amount of energy reflected is less than $100\%$ (A: $53.0\%$, B: $80.0\%$). Data from both simulations A and B seem consistent with the reflection hypothesis, though we observe less absorption of the ion fluctuations than expected. 
One reason may be that this definition of reflection coefficient does not account for the continuous thermal excitation.  
As shown in Fig.~\ref{fig:density_flucs}, the left moving modes in the ion fluctuations are consistent with thermal noise near the sheath edge, whereas the right moving modes have a higher level of fluctuations. 
In contrast, the electron wave power is nearly identical in both the left and right moving modes. 
%\textcolor{red}{Is this because there is still thermal emission that is not accounted for in the definition of the reflection coefficient?}
In the uniform simulation we observed that both ion and electron fluctuations are fully reflected.  This is the expected behavior since all particles reflect from the walls in simulation C.

In summary, the observations presented in Fig.~\ref{fig:density_flucs} are consistent with the hypothesis that instabilities are present when the electron temperature exceeds the predicted threshold for ion-acoustic instability excitation in the presheath, and that some fraction of ion-acoustic fluctuations that impinge on the sheath are reflected back into the plasma, most likely by the electrons. 
Other works also suggest that ion-acoustic waves reflect from sheaths.\cite{ishihara_reflection_1978,bertotti_reflection-absorption_1966,ibrahim_reflection_1984,berumen_analysis_2018}
%However, our results do not represent the only study of ion-acoustic reflection from sheaths as reflection coefficient have been discussed in theoretical detail \cite{ishihara_reflection_1978,bertotti_reflection-absorption_1966,ibrahim_reflection_1984} and propose that reflection is possible.  
Recently, Hood \emph{et al.}~\cite{hood_laser-induced_2020} measured fluctuations in an ion presheath at conditions where ion-acoustic instabilities were expected to be excited near the sheath edge.  The measurements showed evidence of ion-acoustic fluctuations near the sheath edge, but also further into the plasma in the region predicted by linear theory to be stable~\cite{hood_laser-induced_2020}.  Furthermore,  when the biased probe on which the sheath formed was removed, and the measurement repeated, no fluctuations were measured to within the diagnostic resolution. This shows that ion-acoustic fluctuations are due to the presence of the sheath, and suggests that they were reflected from the sheath back into the bulk plasma. Our simulations are consistent with this picture, but the high noise level in PIC simulations makes it difficult to conclusively associate enhanced fluctuations with an instability based on the spectrum alone.

\subsection{Enhanced electron-ion energy exchange rate \label{sec:e_balance}}

As mentioned earlier, ion heating in simulations with electron temperatures above $0.7$ eV seem to be attributable to an increased electron-ion energy exchange rate caused by ion-acoustic instabilities. To test whether this is the case, we evaluate terms in the ion energy evolution equation and find that the term representing instability-enhanced electron-ion scattering is significant in a simulation with $T^s_e$ above the instability threshold (simulation A from Sec.~\ref{sec:instability}). Furthermore, in a simulation with an electron temperature below the threshold (simulation B from Sec.~\ref{sec:instability}) the instability-enhanced term is near zero implying that there is no instability-enhanced heating.

The ion temperature equation is obtained from the second velocity moment of the ion kinetic equation, which in one spatial dimension ($x$) is
\begin{equation}
    \partial_tf_i+v_x\partial_{x}f_i+\frac{e}{m_i}E\partial_{v_x}f_i = C_c+C_n +R^s.
    \label{eq:1dkinetic}
\end{equation}
Here, $C_c$ and $C_n$ are the Coulomb and neutral collision operators and $R^s = \nu^s_if^s_i$ represents the sourcing of ions at a fixed rate $\nu^s_i$ from the distribution $f^s_i$, a stationary Maxwellian with temperature $T^s_i = 0.026$ eV.  
Each term on the right side of Eq.~(\ref{eq:1dkinetic}) represents a potential mechanism for heating.  The source term (representing ionization collisions) can broaden the IVDF in the presheath by introducing new particles at velocities much less than the average ion velocity.  The $C_n$ term represents other ion-neutral collisions, like charge exchange, which can lead to ion heating when a fast ion is effectively slowed down. However, at a pressure of 0.01 mTorr this term is negligible since the ion-neutral mean free path is approximately 100 times longer than the simulation length.  Equally as rare are Coulomb collisions between electrons and ions, represented by $C_c$, since the electron-ion mean free path is also much larger (thousands of meters) than the simulation length. However, if ion-acoustic instabilities are present, they can significantly increase the collision rate between electrons and ions \cite{baalrud_kinetic_2008}. The increased collision rate would allow for more energy exchange between electrons and ions, heating the ions.

To determine the electron-ion energy exchange rate we calculate the $\frac{m_i}{2}v^2$ moment of Eq.~(\ref{eq:1dkinetic}), which can be cast in terms of the ion temperature. In one spatial dimension and three velocity dimensions the ion temperature equation takes the form:
\begin{equation}
    \frac{3}{2}n_iV_i^x\dv{T^x_i}{x} + n_iT^x_i\dv{V_i^x}{x} + \dv{q_i^x}{x} +\pi_i^{xx}\dv{V_i^x}{x} =  Q_c +\frac{3}{2}\nu^s_in_iT^s_i.
    \label{eq:1d_temp_evo}
\end{equation}
Here, $n_i$ is the average ion density, 
\begin{equation}
    q^x_i = \frac{m_i}{2}\int(v_x-V_i^x)\abs{\vb{v}-\vb{V}_i}^2f_i\dd[3]{v}
\end{equation}
is the $x$-component of the heat flux, 
\begin{equation}
    \pi_i^{xx} = m_i\int\qty((v_x-V_i^x)^2-\frac{1}{3}\abs{\vb{v}-\vb{V}_i}^2)f_i\dd[3]{v}
\end{equation}
is the $xx$-component of the stress tensor,
and 
\begin{equation}
    Q_c = \frac{m_i}{2}\int\abs{\vb{v}-\vb{V}_i}^2C_c\dd[3]{v}
\end{equation}
is the energy moment of the Coulomb collision operator. Since $C_n$ is negligible at low pressure its moment is not included here. The second term on the right hand side represents the energy introduced by the ion source.
% that would occur in a warm Tonks-Langmuir model \cite{tonks_general_1929,sheridan_solution_2001}.

The contribution of the electron-ion energy relaxation rate to the ion temperature equation is determined by evaluating each of the terms on the left side of Eq.~(\ref{eq:1d_temp_evo}) using the simulated IVDFs, and subtracting from this $\frac{3}{2} \nu_i^s n_i T_i^s$ to obtain $Q_c$; see Fig.~\ref{fig:Qi_low}.
The individual terms from simulation B have nearly equal magnitude up until the sheath edge where the flow gradient terms dominate. In simulation A we see similar behavior near the center, but the term related to the stress tensor reverses sign and contributes significantly near the sheath edge. Summing the term on the left hand side of Eq.~(\ref{eq:1d_temp_evo}) and subtracting the source term results in the residual $Q_c$ which represents the energy exchange rate between electrons and ions. In simulation B the energy exchange rate is effectively a flat profile across the entire domain and has a comparable magnitude to the source term (horizontal line). This is what is expected in a simulation where ion heating results only from the sourcing of particles throughout the presheath. However, the picture changes dramatically in simulation A where the residual term significantly exceeds the source term throughout the presheath, but most notably near the sheath edge. This supports the idea that the heating observed in simulation A is the result of instability-enhanced energy exchange between electrons and ions, since the residual is only significant in simulation A. 

Although the instability-enhanced energy exchange rate decreases significantly further from the sheath edge, it is still far larger than the source term ($\frac{3}{2}\nu^s_in_iT^s_i\approx2\times 10^{-3} \frac{\textrm{W}}{\textrm{m}^3}$) near the center.  This indicates that the heating we observed near the center of simulation A is also caused by the instabilities and further supports the reflection mechanism discussed in the Sec.~\ref{sec:instability}.

\section{Higher Pressure}

\begin{figure}
    \centering
    \includegraphics[width=.5\textwidth]{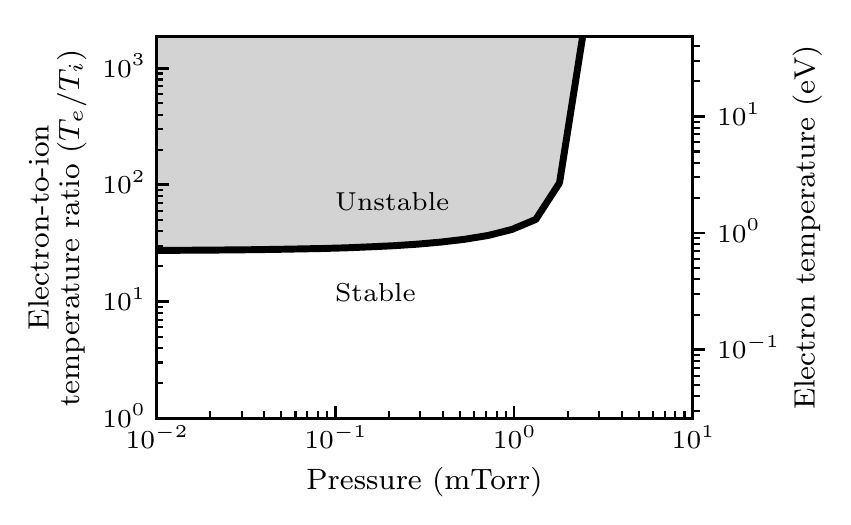}
    \caption{The theoretical temperature ratio threshold as a function of pressure calculated using Eq.~(\ref{eq:bgk_gamma}).}
    \label{fig:theshold_v_pressure}
\end{figure}

Increasing the neutral pressure has two major consequences: increasing the temperature ratio threshold necessary for exciting the instabilities and increasing the heating that results from ion-neutral collisions in the presheath.  
Figure~\ref{fig:theshold_v_pressure} shows an estimate for the threshold temperature ratio as a function of pressure illustrating the former point. 
To study the ion heating due to ion-neutral collisions, with and without the instabilities present, we included simulations at pressures of $0.01$ mTorr, $1$ mTorr and $10$ mTorr with individual simulations taking the same electron temperature values as in Table \ref{tab:Te_ssn}.  We find that the heating from ion-neutral collisions is isotropic, which distinguishes it from instability-enhanced heating, which is anisotropic (in 1D-3V simulations).

\begin{figure}
    \centering
    \includegraphics[width=\linewidth]{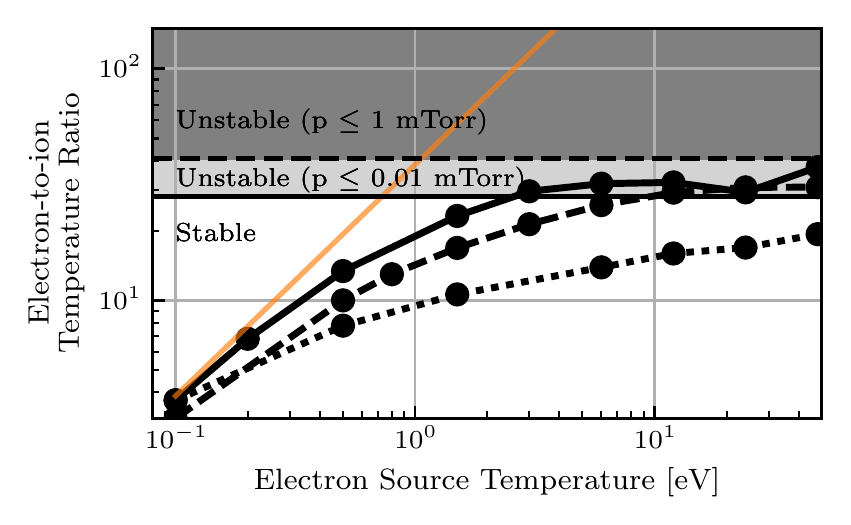}
    \caption{Electron-to-ion temperature ratio at the sheath edge vs. $T^s_e$ for pressures of 0.01 mTorr (solid), 1 mTorr (dashed), and 10 mTorr (dotted). The source temperature ratio is shown in solid orange and the solid horizontal line represents the collisionless (0.01 mTorr) prediction for the threshold $(T_e/T_i)^{\textrm{th}} = 28$ and the dashed line represents the threshold at 1 mTorr $(T_e/T_i)^{\textrm{th}} = 41$.}
    \label{fig:Tr_vpress}
\end{figure}

\begin{figure}
    \centering
    \includegraphics[width=\linewidth]{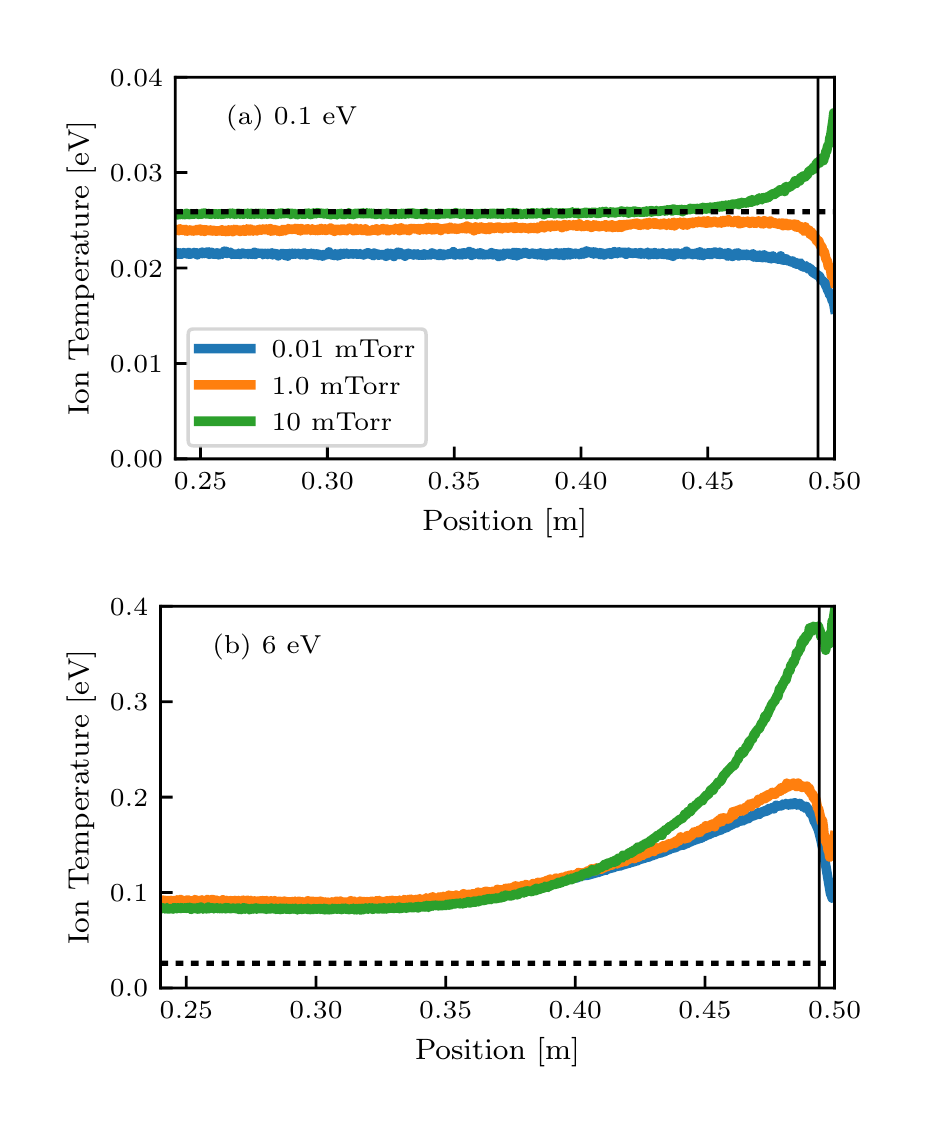}
    \caption{Ion temperature profiles for different pressures at electron temperatures of (a) $0.1$ eV and (b) $6$ eV representing simulations where ion-acoustic instabilities are not present and present only at low enough pressures, respectively. The dashed horizontal line represents the ion source temperature (room temperature).}
    \label{fig:Ti_vpress}
\end{figure}

The effects of ion-neutral collisions can be modeled using the Bhatnagar–Gross–Krook\cite{bhatnagar_model_1954} (BGK) collision operator. It uses a velocity-independent collision frequency which can be estimated from the Phelps database cross sections of $He^{+}$ and $He$ ($\sigma_{i-n}$) \cite{noauthor_phelps_nodate-1,noauthor_phelps_nodate}. Ultimately the BGK model predicts a modification to the growth rate of the form
\begin{equation}
    \label{eq:bgk_gamma}
    \gamma_{\textrm{BGK}} = \gamma - \nu_{i-n}/2,
\end{equation}
where $\gamma$ is the growth rate from Eq.~(\ref{eq:gamma}) and $\nu_{i-n}$ is the ion neutral collision frequency estimated as $\nu_{i-n}\approx \sigma_{i-n}c_sn_{He}$. In Fig.~\ref{fig:theshold_v_pressure} we see that the collisionless prediction of $(T_e/T_i)^{\textrm{th}} = 28$ is returned at low pressures and at $1$ mTorr the temperature ratio threshold is about $(T_e/T_i)^{\textrm{th}} = 41$, which is a relatively minor change corresponding to an electron temperature threshold of about 1.1 eV. 
Finally, above approximately 3 mTorr the threshold temperature ratio becomes so large (roughly 1000) that no ion-acoustic instabilities are expected over the entire range of simulated parameters.

Figure~\ref{fig:Tr_vpress} shows the sheath edge temperature ratio for 3 different pressures. 
The main observation is that higher neutral pressure lowers the observed electron-to-ion temperature ratio for all source electron temperatures considered. 
Although the predicted instability threshold is only slightly higher at 1 mTorr than at 0.01 mTorr, 1 mTorr is apparently a high enough pressure to cause some ion heating due to interaction with neutrals. 
As a consequence, the observed temperature ratio does not reach the predicted threshold value over the range of simulated values. 
At a pressure of 10 mTorr, heating due to ion-neutral collisions is even more significant. 
This is consistent with the expectation that no ion-acoustic instabilities should be present in the 10 mTorr simulation.

Simulations at low electron source temperatures ($T^s_e = 0.1$ eV) show that ion-neutral collisions result in minor ion heating throughout the presheath. 
Specifically, ion temperature profiles from simulations with neutral pressures of 0.01, 1, and 10 mTorr are plotted in Fig.~\ref{fig:Ti_vpress} (a). 
For all pressures, the temperature profile is flat through most of the domain. 
At 0.01 mTorr and 1 mTorr a slight decrease of the ion temperature is observed near the sheath, whereas a slight increase is observed at 10 mTorr. 
%Even in the higher pressure case, ion-neutral heating is weak because the low electron temperature results in a small sound speed, preventing the IVDF from spreading to a high temperature in the presheath. 
In the simulations with electron temperatures of 6 eV we expect the effects of the ion-acoustic instabilities to be present at low pressure (0.01 mTorr and 1 mTorr), and the effects of ion-neutral collisions to become important at higher pressure (1 mTorr and 10 mTorr). 
In each of the simulations shown in Fig.~\ref{fig:Ti_vpress} (b), the temperature in the bulk is approximately the same, and heating is observed near the sheath. 
The heating is largest at the highest neutral pressure. It is unexpected that the ion heating near the center of the domain persists at higher pressures where the instabilities should be damped.

\begin{figure}
    \centering
    \includegraphics[width=\linewidth]{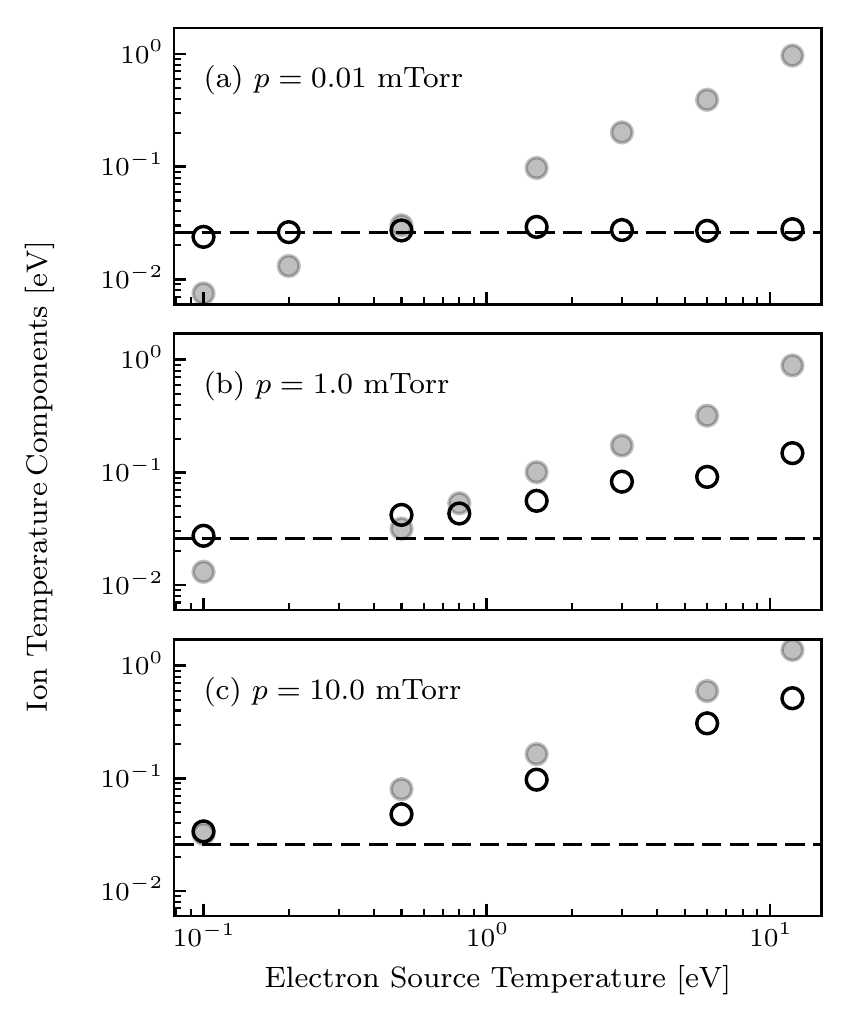}
    \caption{Parallel (solid markers) and perpendicular (open markers) ion temperatures at the sheath edge for different electron temperatures at pressures of 0.01 mTorr, 1 mTorr, and 10 mTorr. At higher pressures fewer simulations were run due to increased computation time.}
    \label{fig:Tperp_vpress}
\end{figure}

\begin{figure*}
\centering
\includegraphics[width=\textwidth]{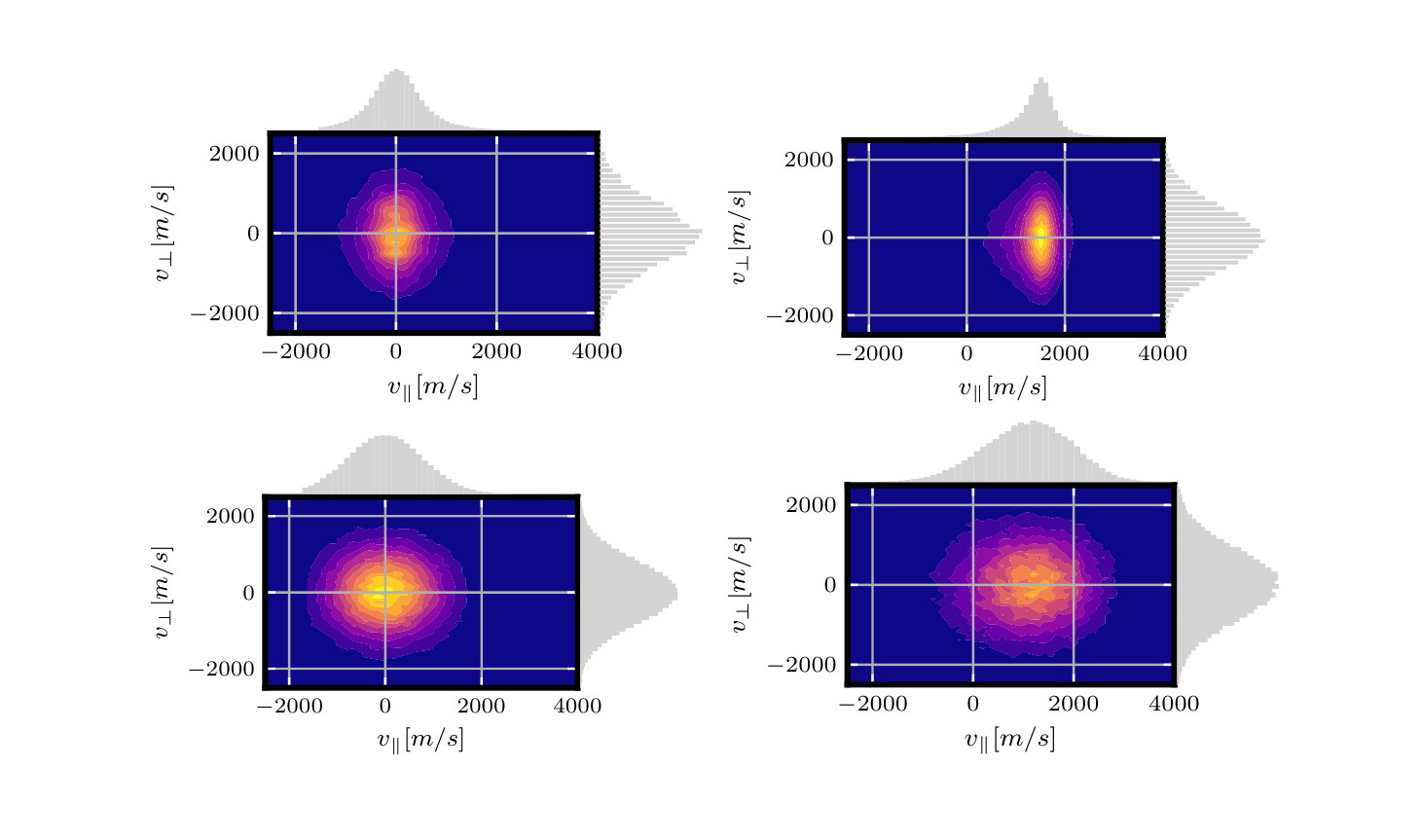}
\caption{IVDFs in the $v_{\perp}$ vs $v_{\parallel}$  ($v_{y}$ vs $v_x$) plane.  The source electron temperature in each simulation was 0.1 eV, which is low enough that no instabilities are expected. Each row shows the IVDF at two different locations from left to right: $0.25$~m, $0.49$~m. The top row represents a collisionless system with $p_n = 0.01$ mTorr and the bottom row a collisional system with $p_n = 10$ mTorr}
\label{fig:IVDFs_vpressure}
\end{figure*}

\begin{figure}
    \centering
    \includegraphics[width=\linewidth]{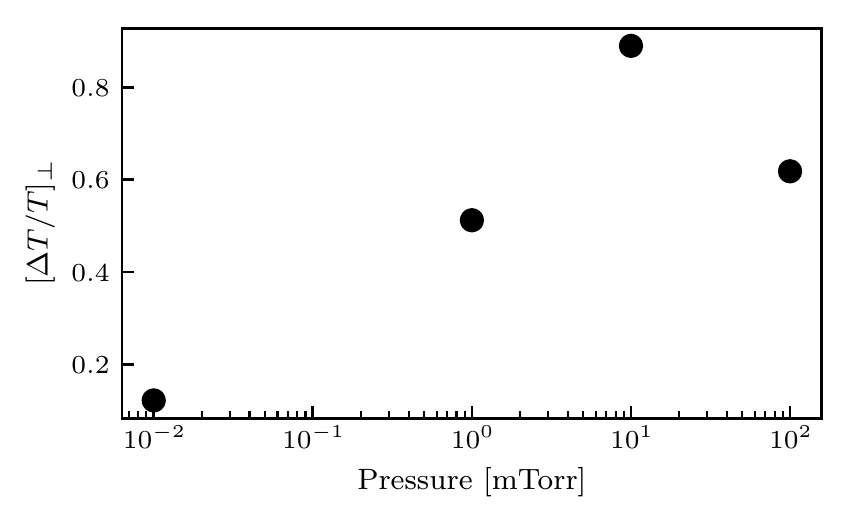}
    \caption{Change in perpendicular ion temperatures across the presheath plotted against pressure for an electron temperature of 0.5 eV.}
    \label{fig:dT_perp}
\end{figure}

It is important to point out that the heating from ion-neutral collisions increases with electron temperature as well as neutral pressure. This is seen in Fig.~\ref{fig:Ti_vpress} where the ion temperature is very close to the source temperature in (a), but much higher in (b). The reason for this is that the sound speed is approximately $\sqrt{60}$ times higher in (b) than (a) since the electron temperature increases by a factor of 60.  Plasmas with a higher sound speed experience more ion-neutral heating since the IVDF spans a wider range of velocities ($0\xrightarrow {}c_s\propto \sqrt{T_e}$) while traversing the presheath.  Ion-neutral collisions drag the IVDF out over these lower velocities, resulting in a wider IVDF in plasmas with higher electron temperatures.  This makes it difficult to differentiate between instability-enhanced heating and heating from ion-neutral collisions since they increase simultaneously.
%However, we observe a significant difference between the two when we look at how each affects the perpendicular IVDF.

%To illustrate this point we estimate the pressure at which the transition occurs 

Furthermore, the two mechanisms are not separated by a pressure regime where neither is significant.  Instead, when the electron temperature is sufficient for instabilities, the dominant heating mechanism immediately transitions from instability-enhanced to collisional with increasing pressure.  This is demonstrated by calculating an estimate of the neutral pressure at which the instability-enhanced energy exchange rate ($Q_c$) becomes equivalent to the ion-neutral energy exchange rate ($Q_n = \frac{m_i}{2}\int\abs{\vb{v}-\vb{V}_i}^2C_n\dd[3]{v}$). We can estimate $Q_c$ from Fig.~\ref{fig:Qi_low} (c) (which assumes $T^s_e = 6$ eV) and so take $Q_c \approx 1$ W/m$^3$.  Here, $Q_n$ was estimated assuming that the ion-neutral collision operator has the same form as the ion source term, but with a different collision frequency and distribution ($C_n = \nu_{i-n}n_i e^{-(\vb{v}-\vb{V}_i)^2/v_{Ti}^2}/(\pi^{3/2}v_{Ti}^3)$).  This distribution represents the steady-state that the IVDF should relax to in the presence of ion-neutral collisions. In this model, the ion-neutral energy exchange rate is
\begin{equation}
    Q_n\ [\textrm{W/m}^3]\approx \frac{3}{2}\nu_{i-n}n_iT_i \approx \frac{3}{2}\sigma_{i-n}c_sn_{He}n_iT_i,
\end{equation}
where the ion-neutral collision frequency was approximated as $\nu_{i-n} \approx \sigma_{i-n}c_sn_{He}$. The neutral density can be estimated for a given temperature (here room temperature) and neutral pressure $p$ (measured in mTorr). The estimate for the ion-neutral energy exchange rate is
\begin{equation}
    Q_n\ [\textrm{W/m}^3]\approx 0.3 p\ [\textrm{mTorr}],
\end{equation}
%  $n_{\textrm{He}}\ [m^{-3}] \approx 3.250 \times 10^{19} p$ [mTorr]
where the electron temperature was taken to be 6 eV and the ion-neutral cross section was estimated as  $\sigma_{i-n} \approx 2\times 10^{-19} \textrm{m}^2$ from the Phelps database \cite{noauthor_phelps_nodate,noauthor_phelps_nodate-1}.
Equating this to the instability enhanced energy exchange rate results in an estimated transition pressure around 3 mTorr.  This implies that at pressures well below this value the instability-enhanced heating is most important, but at intermediate pressures ($\approx 1$ mTorr) both mechanisms will contribute to ion heating. 
This estimate appears to be consistent with the observations, and is further corroborated by considering the ion temperature anisotropy. 

Specifically, elastic ion-neutral collisions lead to a much more isotropic IVDF than the instability-enhanced heating and the transition from anisotropic to isotropic agrees well with the estimated transition of 3 mTorr. This is depicted in Fig.~\ref{fig:Tperp_vpress}, where the perpendicular (open markers) and parallel (solid markers) ion temperature at the sheath edge is plotted as a function of electron temperature for several pressures. Clearly ion-neutral collisions transfer some of the parallel energy the ions gain in the presheath into the perpendicular direction. The simulations also indicate that this transfer in energy is most significant at the highest neutral pressure (10 mTorr), where the heating is nearly isotropic. 
It is nonexistent at the lowest pressure (0.01 mTorr), and has an intermediate value at 1 mTorr. 
This seems consistent with the predicted transition from ion-acoustic instability to ion-neutral collisional heating occurring near 1 mTorr neutral pressure. 

At electron temperatures where we don't expect ion-acoustic instabilities we find that our observation agree well with previous PIC simulations \cite{meige_ion_2007}. 
%that explored the effects of neutral pressure on ion temperatures in the presheath. 
In the simulations of Meige \emph{et al.}, the electron density was modeled using the Boltzmann relation instead of the PIC method and so excluded the excitation of ion acoustic instabilities which require inverse Landau damping to occur~\cite{gurnett_introduction_2017}.  Specifically, we compare their results to our simulations with electron temperatures of  0.1 eV. 

Meige \emph{et al.} report a slightly anisotropic IVDF at low pressure (0.1 mTorr) where the ion temperature varies little across the presheath. This is consistent with our simulations, demonstrated in the top row of Fig.~\ref{fig:IVDFs_vpressure}, where the 2D IVDF at 0.01 mTorr has a perpendicular temperature that is nearly twice as large as the parallel temperature.  
The temperatures in the center (``c'') and sheath edge (``se'') positions are also nearly equal ($T^{\parallel}_c = 0.014$ eV, $T^{\parallel}_{se} = 0.013$ eV, $T^{\perp}_c = 0.026$ eV, $T^{\perp}_{se} = 0.026$ eV). 
The lower ion temperature in the parallel direction is associated with ion energy loss to the walls in this dimension. 
However, at a higher pressures (e.g. 10 mTorr) Meige \emph{et al.} report significant broadening of the parallel IVDF, leading to a more isotropic distribution. Again, we see the same behavior as demonstrated in the bottom row of Fig.~\ref{fig:IVDFs_vpressure}, where the IVDF in the $v_{\perp}$ vs $v_{\parallel}$ plane is nearly circular, with slightly higher temperatures near the sheath edge ($T^{\parallel}_c = 0.024$ eV, $T^{\parallel}_{se} = 0.032$ eV, $T^{\perp}_c = 0.026$ eV, $T^{\perp}_{se} = 0.032$ eV).

In addition to detailing the broadening of the IVDF in the presheath, Meige \emph{et al.} discovered that the ions heat in the perpendicular direction as they transit the presheath and that this effect has a non trivial dependence on the pressure.  Specifically, the perpendicular heating due to ion-neutral scattering increases until about 10 mTorr and then decreases at higher pressure. Fig.~\ref{fig:dT_perp} shows the change in perpendicular temperature of the ions across the presheath ($\qty[\Delta T/T]_{\perp} = \qty(T^{\perp}_i(\textrm{se})-T^{\perp}_i(\textrm{c}))/T^{\perp}_i(\textrm{c})$). Our simulations show similar behavior to those of Meige \emph{et al.}, though with a slightly higher perpendicular heating at all pressures.  Specifically, at 10~mTorr our simulations predict a percent increase of 0.9 while Meige \emph{et al.} predict 0.75. This further supports the fact that our simulations agree well with previous simulations when we chose the electron temperature to be low enough that the ion-acoustic instabilities are not excited.

%=======================================================================
%=================== Summary ===========================================
%=======================================================================

\section{Conclusions}
 
We observe significant ion heating in the presheaths in 1D PIC-DSMC simulations when the electron-to-ion temperature ratio is high enough to excite ion-acoustic instabilities near the sheath edge. 
We observe enhanced ion-acoustic fluctuation levels alongside ion heating when the instabilities are expected. Further analysis of our simulations support the hypothesis that excited waves greatly increase the rate of electron-ion energy equilibration such that the electron-to-ion temperature ratio at the sheath edge is near the threshold value at which instability onsets. 
Wave reflection is observed and is a plausible mechanism allowing for the heating to occur away from the sheath. The heating we observe exceeds the heating expected from inelastic collisions that occur in the presheath at low pressure.  The amount of heating observed increases with the source electron temperature. 
At sufficiently high neutral gas pressure, the instability-enhanced heating mechanism is replaced by collisional heating (here at pressures above 1 mTorr).  However, ion heating is still observed far from the sheath edge, which is unexpected at high pressures. Currently available experimental data are consistent with what we observe in our simulations, but measurements of the ion temperature over a range of electron temperatures will be needed to provide a definitive test. 
In addition, the instability-enhanced heating observed is limited to the parallel direction of our 1D-3V simulations; however this may be an artifact of the 1D simulations since the electric field is confined to the one spatial dimension.  2D or 3D simulations will be necessary to confirm if this is the case and what effect it has on the heating.  The results presented in this work provide another example of how electrostatic instabilities driven by ion flow influence transport in the presheath.~\cite{BaalrudPSST2016}

% \textcolor{blue}{The results presented in this work provide another example of how electrostatic instabilities driven by ion flow influence transport in the presheath. 
% Other recent examples ...}
% Finally, the analysis in this work is similar to earlier work on the ion-ion two stream instability in two ion species plasmas \cite{adrian_influence_2017,baalrud_interaction_2020}. In such plasmas the ion-ion instability resulted in an increased friction force between the two ion species and ultimately locked the difference between ion average velocities to a critical value \cite{kim_ion-neutral_2017}. 

%=======================================================================
%=================== Acknowledgements ==================================
%=======================================================================

\section{Acknowledgments}
The first author would like to thank Brett Scheiner for conversations and tutelage concerning the PIC method. The authors would also like to thank Benjamin Yee, Trevor Lafleur, and David Sirajuddin for their comments on the manuscript.
This material is based upon work supported by the U.S. Department of Energy, Office of Science, Office of Fusion Energy Sciences under contract number DE-NA0003525. This research used resources of the Low Temperature Plasma Research Facility at Sandia National Laboratories, which is a collaborative research facility supported by the U.S. Department of Energy, Office of Science, Office of Fusion Energy Sciences.
The first author was supported by the U.S. Department of Energy, Office of Science, Office of Workforce Development for Teachers and Scientists, Office of Science Graduate Student Research (SCGSR) program. The SCGSR program is administered by the Oak Ridge Institute for Science and Education (ORISE) for the DOE. ORISE is managed by ORAU under contract number DE‐SC0014664. All opinions expressed in this paper are the author’s and do not necessarily reflect the policies and views of DOE, ORAU, or ORISE.

\section{Data Availability}
The data that support the findings of this study are available from the corresponding author upon reasonable request.

%=======================================================================
%=================== References ========================================
%=======================================================================
\section{References}
% \nocite{*}
\bibliography{references}

\end{document}